\documentclass[9pt,twocolumn,twoside]{osajnl}
\usepackage{comment}
\usepackage{soul}
\journal{ol} % Choose journal (ao, aop, josaa, josab, ol, pr)
\setboolean{shortarticle}{true}
\title{In-situ detecting cooperative-target's speed and rotation inertia using structured light}

\author[1,2]{Xiao-Bo Hu}
\author[2]{Bo-Zhao}
\author[2]{Zhi-Han Zhu}
\author[2]{Wei Gao}
\author[2,*]{Carmelo Rosales-Guzm\'an}
\affil[1]{The Higher Educational Key Laboratory for Measuring \& Control Technology and Instrumentations of Heilongjiang Province, Harbin University of Science \& Technology, Harbin 150080, China}
\affil[2]{Wang Da-Heng Collaborative Innovation Center, Heilongjiang Provincial Key Laboratory of Quantum manipulation \& Control, Harbin University of Science and Technology, Harbin 150080, China}

\affil[*]{Corresponding author:carmelorosalesg@hrbust.edu.cn}

\begin{abstract}
Laser remote sensing represents a powerful tool that enables the accurate measurement of the speed of moving targets. Crucially, most sensing techniques are 2-Dimensional and only enable direct determination of the speed along the line of sight. A disadvantage that is very often overcome using two-dimensional techniques that in many cases are hard to implement and expensive. Here we put forward a novel 3-Dimensional technique that enables the direct and simultaneous measurement of both, the speed and the spin rate of cooperative targets. This technique is based on the use of complex vector light beams, whose polarization and spatial degree of freedom are coupled in a non-separable way. We present experimental evidence of our technique by performing a laboratory proof-of-principle experiment. 
\end{abstract}

\begin{document}

\maketitle
The ability to measure the {\it in situ} velocity and spin rate of  targets simultaneously is of great relevance in a wide variety of remote sensing applications such as, space docking, medicine, biology or fluid dynamics, to mention just a few \cite{Yeh1964,BiophysicalJ1972,BiophysicalJ,JReprodFertil}. In general, this 3D-type of motion can be decomposed into two orthogonal motions, one along the line of sight and another in the perpendicular plane. Hence, we can compute each velocity component along each orthogonal direction. Nevertheless, simultaneous measurement of both velocity components is rather cumbersome. Most common techniques rely on indirect two-dimensional measuring schemes, which very often are hard to implement and, in some cases, require extensive post-processing numerical computations \cite{Nagai2009,Nitta1998,BullMathBiol_2011,MeasSciTechnol,DiCaprio:14,PNAS}. This is the case of laser remote sensing, a mature technique which has been developed into compact comercial devices. Most, laser remote sensors are based on the classical non-relativistic Doppler effect, which only allows direct determination of the velocity component along the line of sight \cite{Measures1992}. The transverse component, however, is commonly measured either by using complicated multiple-laser sensing techniques or 3D dynamic imaging systems, which leads to complex and expensive system designs, especially for aerospace applications.

Of great relevance in recent time, was the proposal of a novel technique capable to measure the spin rate using a single laser \cite{Belmonte2012}. This technique relies on the use of beams with a transverse structured phase, which nowadays can be tailored in a wide variety of ways, for example using modern computer-controlled devices, such as Spatial Light Modulators (SLM) \cite{rosales2017shape}. The experimental demonstration of this technique was achieved shortly after its proposal, which even found its way into fluid dynamics by providing with a novel technique to measure directly the vorticity in fluids \cite{SciRep,Lavery2013,Rosales2014OL,Belmonte2015,Ryabtsev2016}. A step further was given with the demonstration of a technique capable to measure the two velocity components involved in a 3D helical motion  \cite{Rosales-Guzman2014}. This technique required the sequential interrogation of the target with two linearly polarized scalar beams, a Gaussian and a helically-twisted vortex beam, to determine the longitudinal and transverse velocity components, respectively. However, the main drawback of this technique is that it can only determine one velocity component at the time, a disadvantage for applications that involve fast movement of the target.

In this work we put forward a novel technique capable to measure {\it in situ} and simultaneously both, the velocity and spin rate of a cooperative target, using a single interrogating beam. For this, we take advantage of so-called vector beams, classically-entangled in their polarization and spatial degrees of freedom (DoF). Vector beams have become topical in recent time due to their unique properties, which have found applications in a wide variety of research fields \cite{Bhebhe2018,Zhan2009,Ndagano2018,Ndagano2017,Berg-Johansen2015,RosalesGuzman2018,Otte2018}. In essence, our technique consists on striking the target with a properly engineered vector beam and use the spatial DoF, encoded on two orthogonal polarizations, to acquire the longitudinal speed and spin rate simultaneously. Hence, under the assumption that the light scattered back from the target preserves its helicity, which is true for scatters that are large with respect to the wavelength \cite{Schwartz2006}, the information stored in each polarization can be unambiguously separated upon detection, for example by interferometric means. 
To illustrate our technique, we simulated the motion of the cooperative target with a rotor mounted on a translation stage. Furthermore, the spatial DoF of the vector beam was generated with an SLM and consisted of a Gaussian and a petal-like structure encoded on left and right circular polarization, respectively. In this way, the Gaussian beam interrogates the target for its longitudinal velocity while the petal-like beam for its spin rate. Crucially, even though in this proof-of-concept we used an SLM to generate the required vector beams, there are several other methods, which can be compact and cheap, such as liquid crystal-based (q-plates) or nanomaterial structures (J-plates)\cite{Marrucci2006,Devlin2017}. Hence, our technique can be incorporated into existing laser remote sensors with minor modifications and mass-produced at very low costs. 

\begin{figure}[tb]
\centering
\includegraphics[width=.47\textwidth]{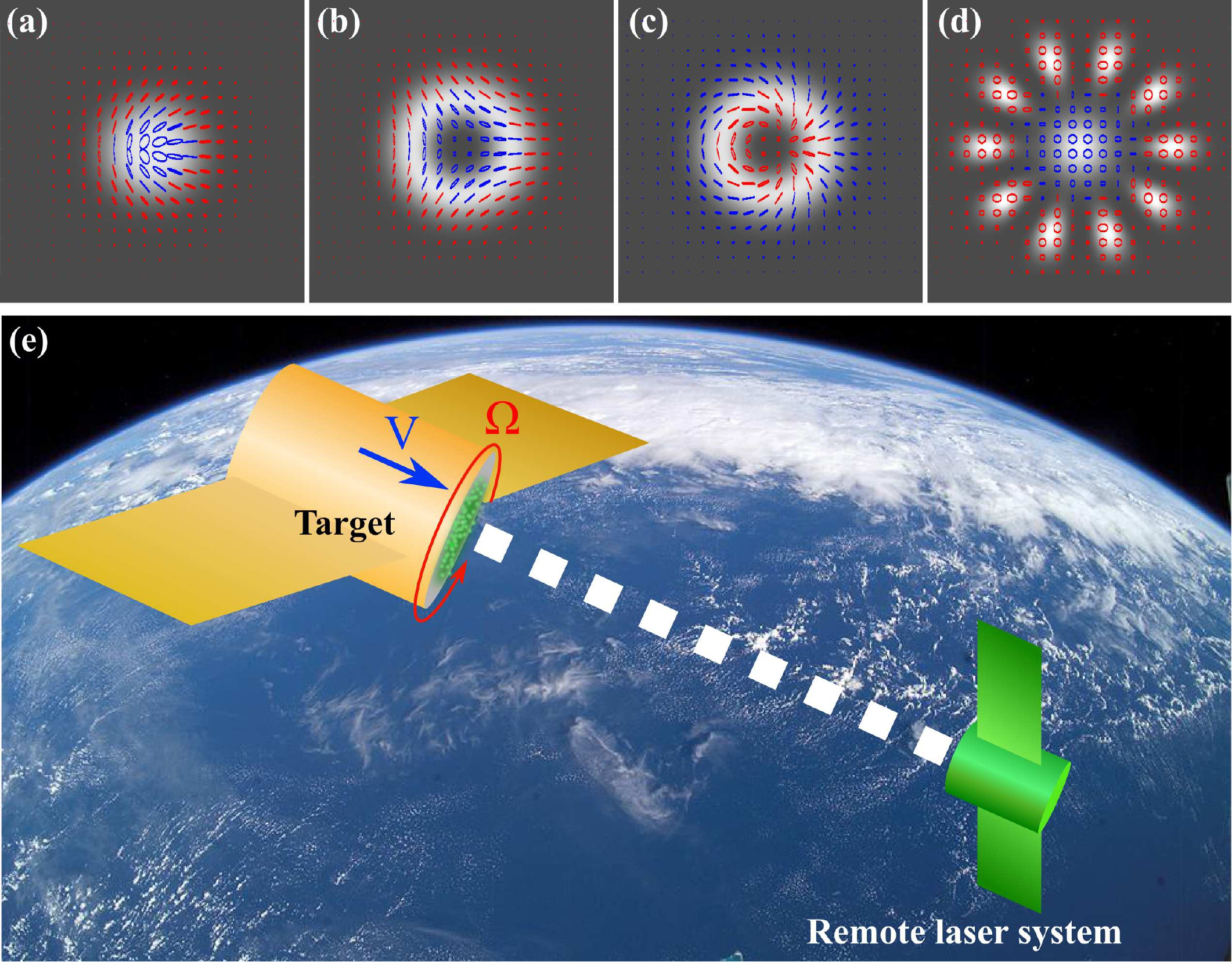}
\caption{Cylindrical vector beams with topological charges: (a) $\ell_1=1$ and $\ell_2=0$; in (b) $\ell_1=-2$ and $\ell_2=-1$; in (c) $\ell_1=1$ and $\ell_2=-2$; in  (d) $\ell_1=5$ and $\ell_2=0$. Here, right elliptical polarization is represented with red color, whereas left elliptical polarization with blue. (e) Schematic representation of a cooperative target being interrogated with a vector beam generated by the remote laser system}
\label{VM}
\end{figure}

Vector beams are commonly generated as non-separable superposition of the spatial and polarization degrees of freedom of light. In a cylindrical coordinate system ($\rho, \phi$) they can be expressed as \cite{Khajavi2016},
\begin{equation}
 U(\rho,\phi)=
\text e^{-\frac{\rho^2}{w^2}}\left\{A\left[{C\text e^{{\text i}\ell_1\phi}+D\text e^{{-\text i}\ell_1\phi}}\right]{\bf\hat e}_{\text R}+B\text e^{{\text i}\ell_2\phi}{\bf\hat e}_{\text L}\right\},
\label{vector}   
\end{equation}
where, ${\bf\hat e}_\text R$ and ${\bf\hat e}_\text L$ are the unitary vectors of the circular polarization basis, $w$ is the half width of the beam, and $\ell$, known as the topological charge, provides the beams with an $\ell\hbar$ amount of orbital angular momentum (OAM) per photon, $\hbar$ being the reduced Plank constant. Further, the coefficients $A=\cos\beta$ and $B=\sin\beta$ ($\beta\in[0,\pi/2]$), enable a continuous variation from fully scalar ($\beta=0, \pi/2$) to fully vector ($\beta=\pi/4$). More over, the terms $C=\cos\theta$ and $D=\sin\theta$ ($\theta\in[0,\pi]$), enables to vary the spatial shape of the right circular polarization term, from a single vortex to a superposition of two vortex beams with opposite topological charges. Equation \ref{vector} gives rise to an infinite set of vector modes with very diverse polarization and spatial distributions, as shown in Fig. \ref{VM}. Here, right and left elliptical polarization are represented by red and blue ellipses, respectively. The case $\theta=0$, $\beta=\pi/4$, which is the most commonly represented in literature, is shown in Fig. \ref{VM}(a) - (c) for the case $\ell_1=1$, $\ell_2=0$; $\ell_1=-2$, $\ell_2=-1$ and $\ell_1=1$, $\ell_2=-2$, respectively. Figure \ref{VM}(d) shows the case $\theta=\pi/4$, $\ell_1=5$ and $\ell_2=0$ featuring a flower-like intensity shape.

In order to determine simultaneously both orthogonal speeds, longitudinal and spin rate, let's recall first that in general, the transverse component can be determined directly by employing a structured light beam with a transverse profile $\Phi({\bf r}_\perp)$  according to \cite{Belmonte2012},
\begin{equation}
    \Delta f_\perp=\frac{1}{2\pi}\nabla_\perp\Phi\cdot {\bf V}({\bf r}_\perp).
    \label{PerpVel}
\end{equation}
Here, $\bf V({\bf r}_\perp)$ and $\nabla_\perp$ represent the velocity and gradient operator, respectively, along the transverse plane. In principle, the transverse structured phase $\Phi({\bf r}_\perp)$ can take any profile but previous knowledge of the target's trajectory enables the appropriate engineering of the same for a further simplification. For example, in the presence of a rotational motion with constant angular velocity $\Omega$, the prove beam can be engineered with a phase of the form $\Phi=\varphi\ell$, with $\ell \neq 0$, to simplify Eq. \ref{PerpVel} to $\Delta f_\perp=\ell\Omega/2\pi$ \cite{SciRep}. Notice that for the same value of $\Omega$, $\Delta f_\perp$ increases proportional to the topological charge $\pm\ell$. The frequency shift $\Delta f_\perp$ can be measured interferometrically by combining the back-scattered light with a reference beam, from which the rotational speed $\Omega$ can be e extracted \cite{SciRep}. 

In our present case, we can use a cylindrical coordinate system to uncouple the motion of the target into two independent motions, a translation along the line of sight, and a rotation along the transverse plane, as illustrated in Fig. \ref{VM}(e). If such an object is illuminated with a vortex beam the back-scattered light is Doppler shifted according to \cite{Belmonte2012},
\begin{equation}
\Delta f=\Delta f_\parallel+\Delta f_\perp= \frac{1}{2\pi}(2kv_z+\ell\Omega),
\label{DopplerTotal}
\end{equation}
where, $k=2\pi/\lambda$ is the magnitude of the wave vector. In the above equation, the first term contains information about the longitudinal velocity, whereas the second term is related to the transverse velocity. Notice whoever that this information is mixed into a single term, making impossible to compute each velocity component independently. One possible way to separate this information is by sequentially illuminating the target with a Gaussian beam followed by a vortex beam $\exp(i\ell\phi)$ \cite{Rosales-Guzman2014}. In this way, under Gaussian illumination only the first term of Eq. \ref{DopplerTotal} is present, whereas, under structured illumination both terms add together into a single frequency shift. From these information, both velocity components can be indirectly determined as $\Delta f_\perp=\Delta f-\Delta f_\parallel$. 

\begin{figure*}[tb]
\centering
\includegraphics[width=\textwidth]{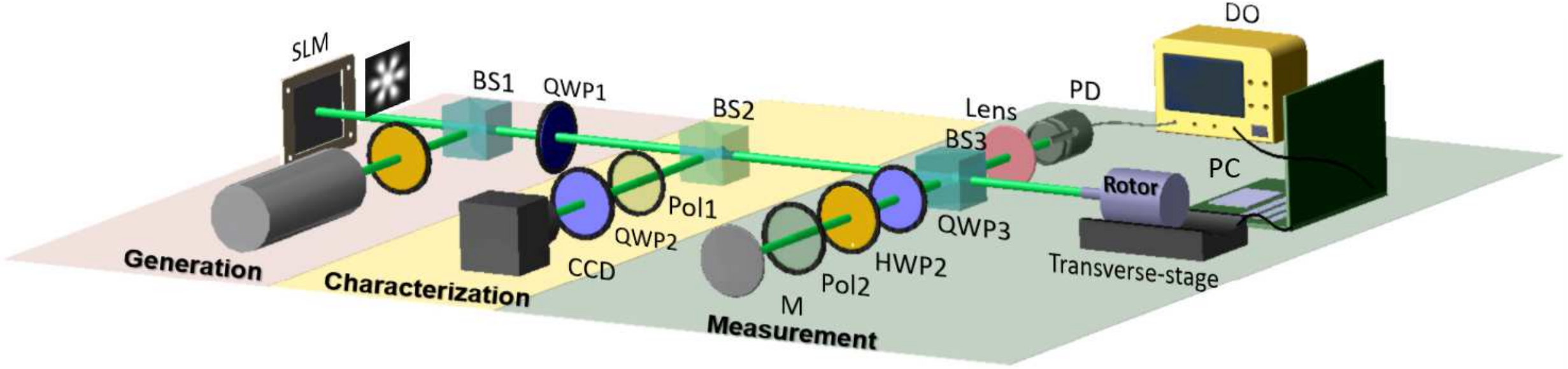}
\caption{Schematic representation of the implemented experimental setup. SLM: Spatial Light Modulator, HWP: Half Wave-Plate, BS: Beam Splitter, QWP: Quarter Wave-Plate, Pol: Linear Polarizer, CCD: Charge-Coupled Device camera, PD: Photodiode, M:Mirror, DO: Digital Oscilloscope.}
\label{setup}
\end{figure*}
Here we propose a novel approach that enables to determine simultaneously and using a single beam of both, the longitudinal velocity and the rotational rate. For this, we illuminate the target with a vector beam of the form,
\begin{equation}
 \begin{array}{ll}
U(\rho,\phi)= \text e^{-\frac{\rho^2}{w^2}}\left\{\frac{1}{\sqrt{2}}A\left[{\text e^{{\text i}\ell\phi} +\text e^{-\text i\ell\phi}}\right]{\bf\hat e}_{\text R}+B{\bf\hat e}_{\text L}\right\},
\label{InField}
\end{array}
\end{equation}
whose intensity and polarization distribution is shown in Fig. \ref{VM}(d), for which, $\theta=\pi/4$, $\ell_2=0$ and $\ell_1=\ell$. The first term is a superposition of two vortex beams carrying opposite topological charges, which enables direct determination of the rotational speed, without the need of interfering it with a reference beam \cite{Lavery2013}. In order to extract the longitudinal velocity component the light scattered back from the target is interfered with a left circularly polarized reference beam. In this way, the reference wave will only interfere with the second term in Eq. \ref{InField}, giving rise to the usual longitudinal Doppler shift, leaving the first  term unaffected. Hence, our approach enables two separate both terms in Eq. \ref{Doppler2} into to independent terms,
\begin{equation}
\Delta f=\Delta f_\perp{\bf\hat{e}}_R+\Delta f_\parallel{\bf\hat{e}}_L= \frac{1}{2\pi}(2\ell\Omega{\bf\hat{e}}_R+2kv_z{\bf\hat{e}}_L),
\label{Doppler2}
\end{equation}
encoded in two orthogonal polarization components.

To demonstrate our technique experimentally, we implemented the setup illustrated schematically in Fig.\ref{setup}, divided in three sections: {\bf generation}, {\bf characterization} and {\bf measurement}. The required vector beams were generated on-axis using a Spatial Light Modulator (SLM), by taking advantage of their polarization dependency \cite{Bhebhe2018}. For this, a linearly polarized Gaussian laser beam ($\lambda=532$nm) is passed through a Half Wave-Plate (HWP), to vary its polarization direction, and directed afterwards to the SLM (Holoeye Pluto 1920x1080, 8 $\mu$m pixel size), using a Beam Splitter (BS1). The SLM is addressed with a superposition hologram encoding two vortex beams with opposite topological charges. To transform the generated beam into the circular polarization basis, a Quarter Wave-Plate (QWP1) is added along the path of the beam. A second beam splitter (BS2), placed after the QWP (see {\bf Characterization}), enables the characterization of the vector beams, prior to their use in the main experiment. Here, the polarization distribution of the vector beam is reconstructed using Stokes polarimetry, for which a series of six intensity measurements were recorded using a linear polarizer (LP) and a QWP (QWP2). Finally, the beam is directed to the measuring section, where a third beam splitter (BS3) is placed to split the beam into a reference and a prove beam. The reference beam is generated using a Michelson configuration (See {\bf Measurement}), where a judicious combination of a QWP, a HWP and a LP enable the removal of the right circular polarization component. The back scattered light is recombined with the reference beam using BS3 and focused, using a lens of focal length $f=30$ mm, onto a photodiode (PD) connected to a Digital Oscilloscope (DO) that enables the recording of a period of the beating intensity signal. The acquired signal is filtered to remove unwanted noise and Fourier transformed to find its frequency content. The required motion of the target was generated using a rotor mounted on a translation stage. Moreover, to enhance the intensity of the back-scattered light, the target was covered with a  retro-reflective painting. It is worth mentioning that all the acquisition process was automated to measure in real time the velocity components of the target.

\begin{figure}[b]
\centering
\includegraphics[width=.46\textwidth]{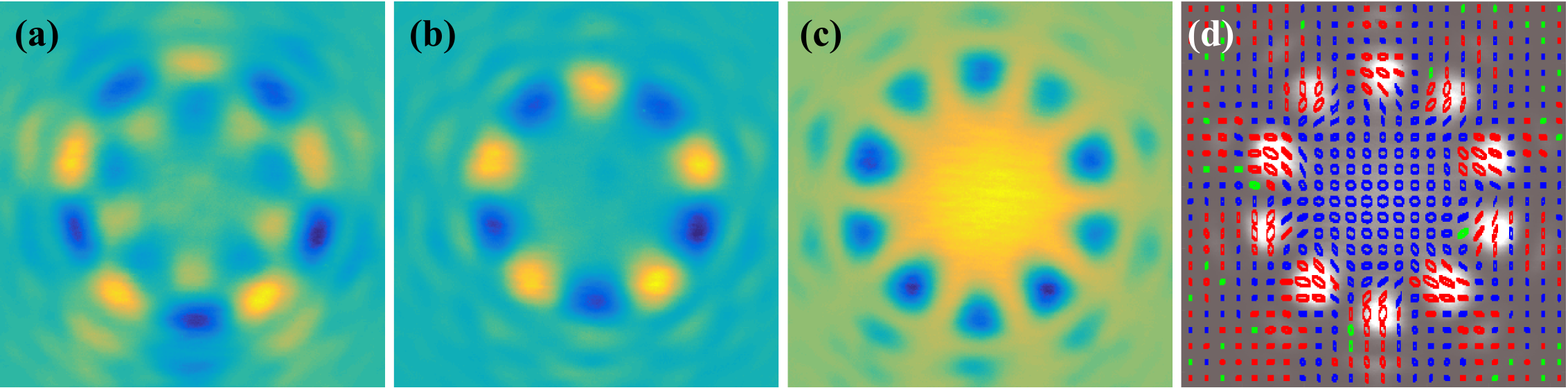}
\caption{Experimental reconstruction of the polarization distribution of one of the employed vector beams. (a) to (c) corresponds to the Stokes parameters $S_1$, $S_2$ and $S_3$, respectively. (d) Reconstructed polarization distribution of the vector beam, overlapped with its intensity distribution.}
\label{Stokes}
\end{figure}

The polarization distribution of the vector beams was reconstructed via Stokes polarimetry. For this, a series of six intensity measurements were recorded, from which the four Stokes parameters defined as,
\begin{equation}
 \begin{array}{ll}
    S_0=I_H+I_V,\qquad
    S_1=I_H-I_V,\\
    S_2=I_D-I_A,\qquad
    S_3=I_R-I_L,
\label{SP}
\end{array}
\end{equation}
were computed. Here, $I_H$, $I_V$, $I_D$, $I_A$, $I_L$ and $I_R$ represent the horizontal, vertical, diagonal, antidiagonal, left circular and right circular, intensity components of the vector beam, respectively. These required intensities were obtained using a LP, in combination with a QWP and recorded with a CCD camera. Figure \ref{Stokes} shows an example of such measurements, whereby the Stokes parameters $S_1$, $S_2$ and $S_3$ are shown in Fig. \ref{Stokes}(a)-(c) respectively, which were used to reconstruct the the vector beam shown in Fig. \ref{Stokes} (d).

\begin{figure}[tb]
\centering
\includegraphics[width=.47\textwidth]{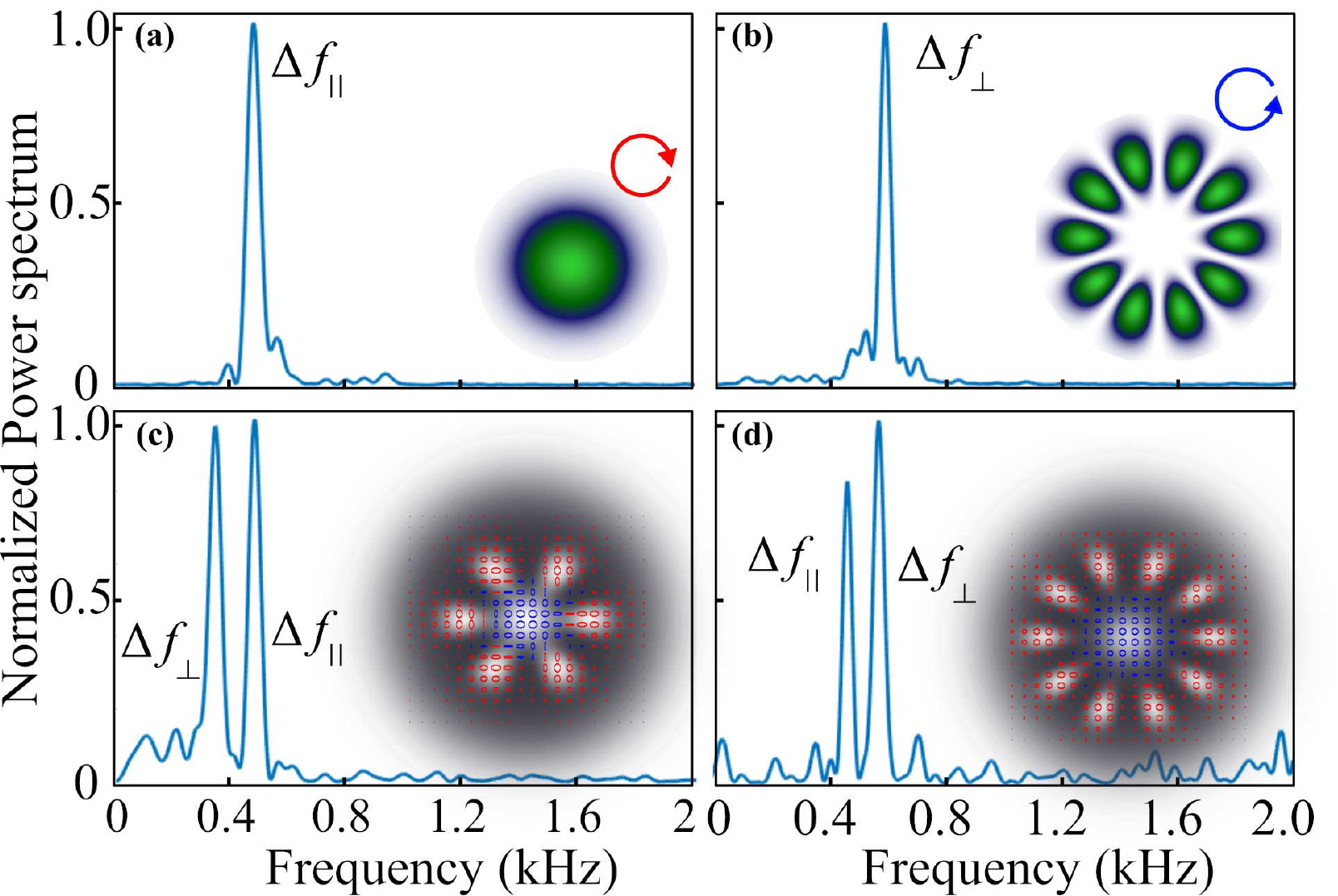}
\caption{Frequency spectrum obtained for a target moving with: (a) pure translation under Gaussian illumination,(b) pure rotation under structured illumination and helical motion under vector beam illumination for (c) $\ell=\pm3$ and (d) $\ell=\pm5$}
\label{Results}
\end{figure}

As a first step to demonstrate our technique, we calibrated both, the translation stage and the rotor. Figure \ref{Results}(a) and \ref{Results}(b) show the frequency spectrum of the target under pure longitudinal and rotational motion, respectively. In both cases the target was interrogated with a vector beam of topological charge $\ell=\pm 5$. First, under pure longitudinal motion the frequency spectrum features a peak at $\Delta f_\parallel=493$Hz, which according to Eq. \ref{Doppler2}, corresponds to the linear speed $v=131.14 \mu$m/s. In comparison to a rotational motion, which gives rise to a frequency peak $\Delta f_\perp=597$Hz, that according to Eq. \ref{Doppler2} yields a rotational speed $\Omega=375.11$ rad/s. Our main results are presented in Fig. \ref{Results}(c) and \ref{Results}(d), where we show two examples of the measured frequency spectrum of the target under both, translation and rotation, clearly evincing the presence of two frequency peaks. Figure \ref{Results}(c) shows the frequency spectrum for the case case $\ell=\pm3$, featuring two peaks at $\Delta f_\parallel=491$Hz and $\Delta f_\perp=358$Hz, which correspond to $v=130.61 \mu$m/s and $\Omega=374.90$rad/s, respectively. Moreover, in Fig. \ref{Results}(d) we show the frequency spectrum for the case $\ell=\pm 5$, for which we get to frequency peaks at $\Delta f_\parallel=483$Hz and $\Delta f_\perp=597$Hz, from which we get $v=128.48 \mu$m/s and $\Omega=375.11$rad/s, respectively.

\begin{figure}[b]
\centering
\includegraphics[width=.4\textwidth]{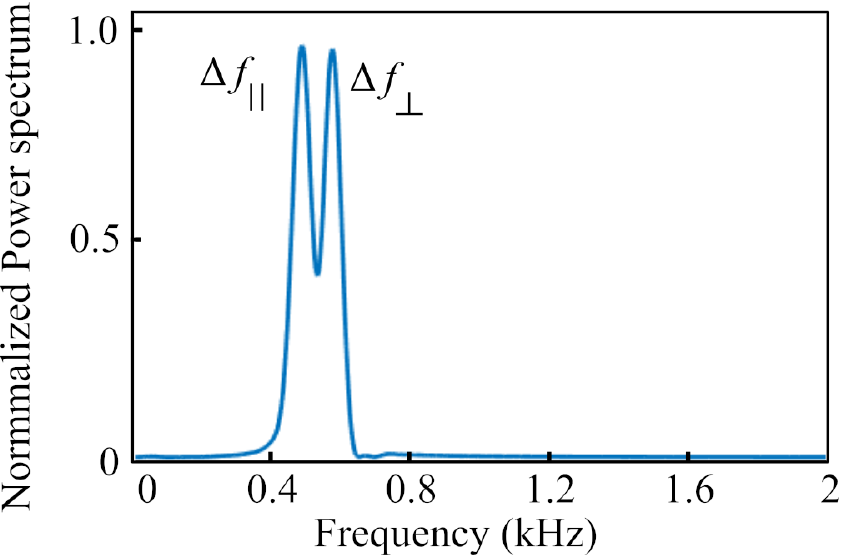}
\caption{Example of a frequency spectrum where the both detected peaks are very close to each other.}
\label{Overlaped}
\end{figure}

It is easy to imagine the highly probable scenario in which, both motions can produce overlapped or very close frequency peaks. Figure \ref{Overlaped} is an experimental example of such case, where the linear and angular speeds correspond to $v=134 \mu$m/s and $\Omega=371$rad/s, respectively, which results in the frequencies, $\Delta f_\parallel=504$Hz and $\Delta f_\perp=591$Hz, respectively. This might seem a drawback of our technique, however, we still have one degree of freedom that plays in our favor, the topological charge of the superposition beam. 

\begin{figure}[tb]
\centering
\includegraphics[width=.45\textwidth]{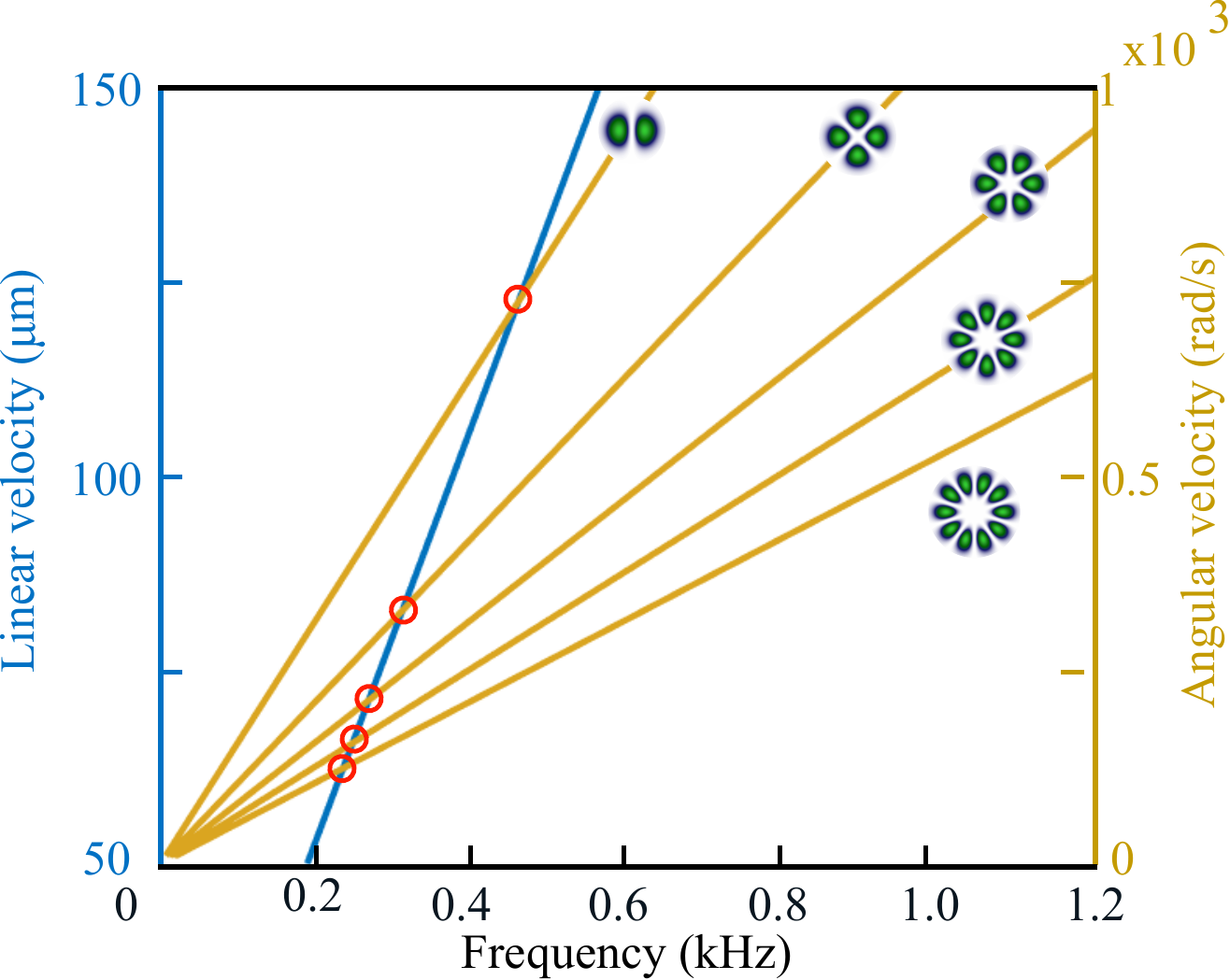}
\caption{Frequency shift (horizontal axis) as function of the linear velocity (left vertical axis) and angular velocity (right vertical axis) illustrating that certain combinations of linear and angular will yield the same frequency peak (red circles) for given $\ell$ values.}
\label{Comparisson}
\end{figure}

Figure \ref{Comparisson} shows a theoretical plot of the frequency shift (horizontal axis) as function of the longitudinal velocity (left vertical axis) compared to the angular velocity for the cases $\ell=\pm5, \pm4, \pm3, \pm2, \pm1$ (right vertical axis). This plot illustrates that for certain longitudinal velocities, the frequency peak will be the same as for rotational velocities measured with certain values of $\ell$, as indicated by the red circles. Take for example the case of a target whose linear and angular velocity are $v\sim62\mu$m/s and $\Omega\sim147$rad/s, it is easy to show, using Eq. \ref{Doppler2}, that the longitudinal frequency peak will be at $\Delta f_\parallel\sim235$Hz. If this target is illuminated with a vector beam, for which $\ell=\pm5$, the measured frequency peak will be very similar. Hence, we would not be able to measure both velocity components, nevertheless, we can always change the topological charge, lets say to $\ell=\pm2$ which will produce the frequency $\Delta f_\perp\sim93$Hz, which now is far away from the longitudinal frequency shift. Hence in the worse case scenario, where both velocity components give rise to a similar frequency peak, a simple change of the topological charge will allow the rotational frequency shift to move away from the longitudinal one. 

%\section{Conclusions}
In short, we have proposed a novel technique that enables the simultaneous determination of both, the longitudinal velocity and rotation rate of cooperative targets. This technique relies on the use of vector beams formed by the non-separable superposition of the spatial and polarization degree of freedom. In this way, while one spatial degree of freedom inquires the target about its longitudinal velocity, the other inquires it about its rotation rate. Given that each spatial shape is encoded on two orthogonal polarizations, each velocity information can be unambiguously measured upon detection. For this, the assumption that back-scattered light preserves its helicity is crucial, an assumption that requires the size of the scatters to be larger with respect to the wavelength of the interrogating beam. Even though here we only presented a proof-of-principle experiment where an SLM was used to generate the required vector beam, there exist other means to generate vector beams, which can be both, compact and inexpensive. Hence, our technique can be easily incorporated into existing laser remote devices, which at the moment can only measure the longitudinal velocity, to produce a device with the capability to measure also the rotation or spin rate of remote targets.

\section*{Funding Information}
 National Natural Science Foundation of China (NSFC) (11574065)

\end{document}